\documentclass[prd,onecolumn,amsmath,amssymb,floatfix,nofootinbib,12pt]{revtex4}
\pdfoutput=1

\usepackage{amssymb}
\usepackage{amsmath}
\usepackage{feynmp}
\usepackage{slashed}
\usepackage{caption}
\usepackage{subfig}
\usepackage{braket}
\usepackage{graphicx}
\usepackage{graphics}
\usepackage{xcolor}
\usepackage{dcolumn}
\usepackage{color}
\usepackage{rotate}
\usepackage{fancyhdr}
\usepackage{indentfirst}
\usepackage{umoline}
\usepackage{hyperref}
\hypersetup{
    colorlinks=true,
    linkcolor=black,
    filecolor=black,
    urlcolor=black,
}

\begin{document}
\title{Measuring CP violation in $b\to c\tau^-\bar{\nu}_\tau$ using excited charm mesons}

\author{Daniel Aloni$^{1a}$, Yuval Grossman$^{2b}$ and Abner Soffer$^{3c}$}
\affiliation{$^1$Department of Particle Physics and Astrophysics, Weizmann Institute of Science, Rehovot, Israel 7610001 \\
$^2$Department of Physics, LEPP, Cornell University, Ithaca, NY 14853\\
$^3$Raymond and Beverly Sackler School of Physics and Astronomy, Tel Aviv University, Tel Aviv,
Israel }
\email{$^a$daniel.aloni@weizmann.ac.il, $^b$yg73@cornell.edu,	$^c$asoffer@tau.ac.il}

\begin{abstract}
\noindent
There is growing evidence for deviation from the standard model
predictions in the ratios between semi-tauonic and semi-leptonic $B$
decays, known as the $R(D^{(*)})$ puzzle. If the source of this
non-universality is new physics, it is natural to assume that it also
breaks CP symmetry. In this paper we study the possibility of
measuring CP violation in semi-tauonic $B$ decays, exploiting
interference between excited charm mesons. Given the current values of
$R(D^{(*)})$, we find that our proposed CP-violation observable could
be as large as about 10\%.  We discuss the experimental advantages of
our method and propose carrying it out at Belle~II and LHCb.

\end{abstract}

\maketitle

\section{Introduction}
Within the Standard Model (SM) of particle physics, the electroweak
(EW) interactions obey flavor symmetry and hence exhibit lepton
flavor universality (LFU). Observation of LFU breaking beyond that of
the Yukawa interactions would be a clear sign of physics beyond the
SM.
In recent years, there have been accumulating experimental indications
for possible LFU violation in the ratios of branching fractions
\begin{align}
R(D^{(*)}) \equiv \frac{BR(\bar B\to D^{(*)}\tau^-\bar{\nu}_\tau)}{BR(\bar B\to D^{(*)}\ell^-\bar{\nu}_\ell)}~,
\end{align}
where $\ell$ denotes an electron or muon.
An average of BABAR~\cite{Lees:2012xj,Lees:2013uzd},
Belle~\cite{Huschle:2015rga,Sato:2016svk,Hirose:2016wfn} 
and LHCb~\cite{Aaij:2015yra,Aaij:2017uff} measurements, calculated by
the Heavy Flavor Averaging Group~\cite{Amhis:2016xyh}, yields
\begin{align}
R(D) = 0.407 \pm 0.046~,
\quad
R(D^*) = 0.304 \pm 0.015~,
\label{eq:RD}
\end{align}
with a correlation coefficient of $-0.2$ between the $R(D)$ and
$R(D^*)$ measurements. 
SM predictions have been calculated in
Refs.~\cite{Lattice:2015rga,Na:2015kha,Aoki:2016frl,Bigi:2016mdz,Fajfer:2012vx}
(for consistency, we follow the predictions considered in
Ref.~\cite{Amhis:2016xyh}),
\begin{align}
	& R_{SM}(D) = 0.299 \pm 0.011 ~\text{\cite{Lattice:2015rga}}~,\quad
	R_{SM}(D) = 0.300 \pm 0.008 ~\text{\cite{Na:2015kha}}~,\\
	& R_{SM}(D^*) = 0.252 \pm 0.003 ~\text{\cite{Fajfer:2012vx}}	~.
\end{align}
The combination of these results deviates by  $4.1\sigma$ from the
SM~\cite{Amhis:2016xyh}.
More recent calculations of $R(D^*)$~\cite{Bernlochner:2017jka,Bigi:2017jbd,Jaiswal:2017rve} reduce this tension somewhat, but do not solve the puzzle.

Another $b\to c\tau\bar\nu$ ratio was recently
measured by LHCb~\cite{Aaij:2017tyk},
\begin{align}
R(J/\psi) = \frac{BR(B_c^-\to J/\psi\tau^-\bar{\nu}_\tau)}{BR(B_c^-\to J/\psi\mu\bar{\nu})} = 0.71 \pm 0.25~.
\end{align}
Although the SM predictions are in the range $R(J/\psi) = 0.25 -0.28$,
the absence of systematic estimation of the uncertainty and lattice
calculations make it debatable whether this measurement increases
the tension with respect to the SM.

The $R(D^{(*)})$ anomaly is puzzling and has received a great deal of
attention. Future measurements, mostly by LHCb and Belle~II, will
greatly reduce the experimental uncertainties. If the disagreement
with the SM becomes significant, it will constitute a clear
signal of physics beyond the SM.
New physics (NP) explanations for this puzzle have been widely
discussed in the literature, where the most popular framework is that
of effective field theory (EFT) with new dimension-six operators that
enhance the taunic decays by about $30\%$ (for a review, see,
{\it e.g.},~\cite{Blanke:2017qan} and references therein).  In order to
explain the enhancement of the central value, all NP solutions
introduce hard breaking of lepton flavor symmetry.

A priori, there is no reason for NP models that solve the $R(D^{(*)})$
puzzle and break lepton flavor symmetry to not break CP at
$\mathcal{O}(1)$ as well. 
Since the SM predicts unobservably small CP violation (CPV) in semileptonic $B$
decays, looking for CPV in $\bar B\to D^{(*)}\tau^-\bar{\nu}_\tau$
can be a clean way to probe physics beyond the SM.

A naive observable of such CP violation is a direct asymmetry in $\bar B\to D^{(*)}\tau^-\nu$ transitions, $i.e.$
\begin{align}
\mathcal{A}_{CP}(\bar B\to D^{(*)}\tau^-\bar{\nu}_\tau) = \frac{BR(\bar B\to D^{(*)}\tau^-\bar{\nu}_\tau) - BR({B}\to \bar{D}^{(*)}\tau^+ \nu_\tau)}{BR(\bar B\to D^{(*)}\tau^-\bar{\nu}_\tau) + BR({B}\to \bar{D}^{(*)} \tau^+ \nu_\tau)}~.
\end{align} 
However, even if there is a NP amplitude with a new weak ({\it i.e.},
CP-violating) phase, this asymmetry is very small due to the absence
of a significant strong ({\it i.e.,} CP conserving) phase between the
interfering amplitudes in this process,

The object of this paper is to introduce and explore a new observable 
that incorporates strong phases, and thus is sensitive to CP violation in models that break lepton universality in $b\to c \tau \nu$ transitions.
Other CPV observables have been suggested in
Refs.~\cite{Duraisamy:2013kcw,Hagiwara:2014tsa}. The main idea in both
cases was to use four-body decay kinematics to construct a triple
product, thus avoiding the need for an explicit strong phase.  To
obtain a four-body decay from $\bar{B}\to
D^{(*)}\tau^-\bar{\nu}_\tau$, one can utilize the subsequent decay of
the $D^*$ into $D\pi$~\cite{Duraisamy:2013kcw}, or the decay of the
$\tau^-$~\cite{Hagiwara:2014tsa}.  
Construction of such observables requires knowing the momentum vectors
of the $\tau^-$ and the $\bar\nu_\tau$ in the $\bar{B}$ rest
frame~\cite{Duraisamy:2013kcw}, or is limited to use of semihadronic
$\tau^-$ decays~\cite{Hagiwara:2014tsa}.

Here we discuss an alternative that is applicable for both leptonic
and semihadronic $\tau^-$ decays. Furthermore, it does not require
measurement of angular variables, although does benefit from even
partial angular information that is experimentally obtainable. 
Our suggestion is to exploit interference between excited
charm mesons.
As can be easily understood in the Breit Wigner
approximation, interference between overlapping resonances gives
rise to strong phases with known phase-space dependence.

The paper is organized as follows: In Sec.~\ref{sec:CPintro} we lay
out our formalism and explain the basic mechanism for generating the
strong phase difference.  In Sec.~\ref{sec:what to measure} we
describe the asymmetry observable we suggest to measure.  In
Sec.~\ref{sec:Simplified Model} we construct a simplified model to
illustrate our method.  Sec.~\ref{sec:RealExperiment} is dedicated to
a discussion of differences between our simplified model and a
realistic experiment. We conclude in Sec.~\ref{sec:conclusion}.

\section{Formalism and basic mechanism}\label{sec:CPintro}

It is well known that observation of a CP asymmetry requires at least
two interfering amplitudes with different weak and strong phases.  Any
new physics that modifies $b\to c \tau \nu$ transitions and breaks CP
naturally provides a second amplitude with a weak-phase difference
relative to the SM amplitude. However, the source of a strong phase is
less trivial.

In our proposal, the strong-phase difference arises from overlapping excited
charm-meson resonances. We consider the decays $\bar{B} \to
D^{**}\tau^-\bar{\nu}_\tau$, where $D^{**}$ is a generic name for the first
four excited charm mesons, $D_0^*$, $D_1^*$, $D_1$, and $D_2^*$.  
The parameters of these states and some of their allowed decays are
listed in Table~\ref{tab:D**}.
The intermediate states $D_0^*$ and $D_2^*$, as well as their
interference, are selected by reconstructing the decay
$D^{**}\to D\pi$. Similarly, the decay $D^{**} \to D^*\pi$ 
selects the states $D_1^*$, $D_1$, and $D_2^*$. 
Generally, the $D_1$ and $D_2^*$ states are easier to study
experimentally, since their widths are smaller. 

Ref.~\cite{Aubert:2008ea} shows a BABAR study of the semileptonic
decays $\bar B \to D^{**}\ell^-\bar{\nu}_\ell$ with $D^{**} \to
D^{(*)}\pi$. The integrated luminosity of Belle~II will be more than
100 times larger, allowing precision measurements of the properties of
the $D^{**}$ states, as well as the $\bar B \to D^{**}$ form factors
needed for interpretation of the $\bar B \to
D^{**}\tau^-\bar{\nu}_\tau$ results. Similar measurements can be
performed at LHCb. Additional studies of the $D^{**}$ states can be
performed with $\bar B\to D^{**}\pi^-$, and in some cases also with
inclusive $D^{**}$ production. We refer to these measurements as
control studies, and note that similar studies were performed
with $\bar B \to D^{(*)}\ell^-\bar{\nu}_\ell$ and $\bar B \to
D^{**}\ell^-\bar{\nu}_\ell$ as part of the measurements of
$R(D^{(*)})$~\cite{Lees:2012xj,Lees:2013uzd,Huschle:2015rga,Sato:2016svk,Hirose:2016wfn,Aaij:2015yra,Aaij:2017uff}. Such studies are also necessary for a $R(D^{**})$ measurement predicted in~\cite{Biancofiore:2013ki,Bernlochner:2017jxt}.

\begin{table}[t]
	\centering
	\begin{tabular}{| c | c | c | c | c |}
		\hline
		Particle & $J^P$ & $M$~(MeV) & $\Gamma$~(MeV) & Decay modes\\
		\hline
		$D_0^*$ & $0^+$ & 2349 & 236 & $D\pi$\\
		$D_1^*$ & $1^+$ & 2427 & 384 & $D^*\pi$\\
		\hline
		$D_1$ & $1^+$ & 2421 & 31 & $D^*\pi$\\
		$D_2^*$ & $2^+$ & 2461 & 47 & $D^*\pi,~D\pi$\\
		\hline
	\end{tabular}	
	\caption{The spin, parity, mass, width, and decay modes of interest
          of the $D^{**}$ mesons~\cite{Bernlochner:2017jxt}
        }
	\label{tab:D**}
\end{table}

In developing our formalism, we make three simplifying assumptions.
\begin{enumerate}
\item The nonresonant $D^{(*)}\pi$ contribution to the ${\bar B} \to D^{(*)}
\pi \, \tau^- {\bar \nu}_\tau$ decay is relatively small over the
narrow $D^{(*)}\pi$ invariant-mass range of interest. While this
contribution should be studied within an experimental analysis, it can
be safely ignored for the purpose of the current
discussion. Therefore, we write the amplitude for this decay as
a sum over the intermediate $D^{**}$ resonances denoted by the index
$i$:
\begin{align}\label{eq:Full amplitude}
\mathcal{A}\equiv\mathcal{A} (\bar{B}\to D^{(*)}\pi\tau^-\bar{\nu}_\tau) & = \sum_i \mathcal{A}(\bar{B}\to D_i^{**}(\to D^{(*)}\pi)\tau^-\bar{\nu}_\tau)~.
\end{align}

\item  We use the narrow-width approximation for the $D^{**}$ mesons.
Then the amplitude for the state $D_i^{**}$ is 
\begin{align} \label{eq:Narrow width}
\mathcal{A}(\bar{B}\to D_i^{**}(\to D^{(*)}\pi)\tau^-\bar{\nu}_\tau)
& = \sum_\lambda
\frac{i \mathcal{A}(\bar{B}\to D_i^{**}(\lambda)\tau^-\bar{\nu}_\tau) \mathcal{A}(D^{**}_i(\lambda)\to D^{(*)}\pi)}{{m^2_{{D^{(*)}\pi}}}-M_{D^{**}_i}^2 +i\Gamma_{D^{**}_i} M_{D^{**}_i}} ~,
\end{align}
where $m^2_{{D^{(*)}\pi}}$ is the invariant mass of the $D^{(*)}\pi$
system, $M_{D^{**}_i}^2$ and $\Gamma_{D^{**}_i}$ are the mass
and width of the intermediate $D^{**}_i$ resonance, $\lambda$ indicates the
helicity of the $D_i^{**}$, and $\mathcal{A}(D^{**}_i(\lambda)\to
  D^{(*)}\pi)$ is the
  $D^{**}_i$ decay amplitude.

\item We further assume that there are no NP contributions in
  $\mathcal{A}(D^{**}_i(\lambda)\to D^{(*)}\pi)$, and that there is
  one NP ${\bar B} \to D^{**} \tau^- {\bar \nu}_\tau$ amplitude with a
  new weak phase $\varphi^{\rm NP}$.  Therefore, we parameterize the
  total ${\bar B} \to D^{**} \tau^- {\bar \nu}_\tau$ amplitude as
\begin{align}\label{eq:B to D** tau nu amplitudes}
\mathcal{A}(\bar{B}\to D_i^{**}(\lambda)\tau^-\bar{\nu}_\tau) & = 
 r_i^{\rm SM}\, e^{i \delta_i^{\rm SM}} + 
 r_{i}^{\rm NP} \, e^{i ( \varphi^{\rm NP} + \delta_{i}^{\rm NP})}~.
\end{align}
Here
\begin{align}
r_i^{\rm SM} &= | C^{\rm SM}| \, |\langle D^{** \, +}_i \tau^- {\bar \nu}_\tau | {\cal O}^{\rm SM} | {\bar B}^0 \rangle |  ~, \label{eq:B to D** tau nu amplitudes parametrized SM}\\
r_{i}^{\rm NP} &= | C^{\rm NP}| \, |\langle D^{** \,
                        +}_i \tau^- {\bar \nu}_\tau |{\cal
                        O}^{\rm NP} | {\bar B}^0 \rangle
                        | \label{eq:B to D** tau nu amplitudes
                        parametrized NP} ~, \\
 \varphi^{\rm NP}&=\arg(C^{\rm NP}), \\
\delta_i^{\rm SM} &= \arg(\langle D^{** \, +}_i \tau^- {\bar \nu}_\tau
|{\cal O}^{\rm SM} | {\bar B}^0 \rangle), \\
\delta_i^{\rm NP} &= \arg(\langle D^{** \, +}_i \tau^- {\bar \nu}_\tau
|{\cal O}^{\rm NP} | {\bar B}^0 \rangle),
\end{align}
where
${\cal O}^{\rm SM}$ and ${\cal O}^{\rm NP}$ are the SM and NP
operators contributing to the transition, respectively, 
and $C^{\rm SM}$ and $C^{\rm NP}$ are the corresponding Wilson coefficients.
We neglect the tiny CP violation in the SM amplitude, and by
redefinition of the states we set the SM weak phase to be $\varphi^{\rm SM}=0$.
\end{enumerate}

In principle, the strong phases $\delta_i^{\rm SM}$ and $\delta_i^{\rm NP}$ 
depend on the kinematics of the event. This dependence is expected to
be small, and we neglect it at that stage.
Up to this
small phase-space dependence, we redefine the states so as to set
$\delta_i^{SM}=0$. 
Furthermore, the strong phases $\delta_i^{\rm SM}$ and $\delta_i^{\rm
  NP}$ are equal in the heavy quark symmetry
limit~\cite{Isgur:1989vq}. Thus, we cannot count on their difference
to be large enough to make it possible to probe CP
violation. Therefore, we adopt a conservative and simplifying
approach, setting all these strong phases to zero. We elaborate on
this in Sec.~\ref{sec:theory-cons}.

A known and large relative strong phase arises from interference
between different overlapping $D^{**}$ meson amplitudes in
Eq.~\eqref{eq:Full amplitude}, particularly in the kinematic region $
m^2_{D^{^{(*)}} \pi} \sim M_{D^{**}}^2$.  
This is the source of strong-phase difference in our proposal. Using
it to generate a sizable CP asymmetry requires that $\bar B \to
D^{**}\tau^-\bar{\nu}_\tau$ amplitudes involving different $D^{**}$
resonances also have different weak phases.
As can be seen from Eqs.~(\ref{eq:B to D** tau nu
  amplitudes}) 
through (\ref{eq:B to D** tau nu amplitudes parametrized NP}), such a weak
phase difference arises only if 
\begin{align}
{r_i^{\rm NP} \over r_i^{\rm SM}} \ne {r_j^{\rm NP} \over r_j^{\rm SM}},
\label{eq:unequal-rates}
\end{align}
{\it i.e.}, if the interfering resonances have different sensitivities
to the NP operator relative to the SM operator.  This happens only if
the resonances have different spins
and the SM and NP operators have different Dirac structures,
{\it i.e.}, $\mathcal{O}_{NP} \ne \mathcal{O}_{SM}$

We emphasize that in the case $\mathcal{O}_{NP} = \mathcal{O}_{SM}$,
even if $\varphi^{\rm NP} \neq \varphi^{\rm SM}$, there is no relative
weak phase between amplitudes involving different $D^{**}$ mesons, and
hence no CP asymmetry. Moreover, in this case, the $\tau$ angular
distributions originating from the SM and NP operators are identical,
so that the previously proposed
methods~\cite{Duraisamy:2013kcw,Hagiwara:2014tsa} also become
insensitive to CPV.

\section{Observable CP asymmetry}\label{sec:what to measure}

A CP asymmetry is obtained by comparing the rate coming
from Eq.~\eqref{eq:Full amplitude} to its CP conjugate,
\begin{align}\label{eq:general-asym}
\mathcal{A}_{\rm CP} = \frac{\int d\Phi \left(\left|\bar{\mathcal{A}}\right|^2-\left|\mathcal{A}\right|^2\right)}{\int d\Phi \left(\left|\bar{\mathcal{A}}\right|^2+\left|\mathcal{A}\right|^2\right)}~,
\end{align}
where $\int d\Phi$ stands for partial phase-space integration, which
is the main issue of this section. 
A four-body decay, such as $\bar{B}\to D^{**}(\to D^{(*)}\pi)
\tau^-\bar{\nu}_\tau$, depends on five kinematical variables. We
choose these to be
\begin{itemize}
	\item $q^2$ - the invariant mass of the $\tau^-\bar\nu_\tau$ system,
	\item $m_{D^{(*)}\pi}$ - the invariant mass of the
          $D^{(*)}\pi$ system,
	\item $\theta_\tau$ - the angle between the $\tau$ momentum and
          the direction opposite the $\bar{B}$ momentum in the
          $\tau^-\bar\nu_\tau$ rest frame, 
	\item $\theta_D$ - the angle between the $D^{(*)}$ momentum
          and the direction opposite the $\bar B$ momentum
          in the $D^{(*)}\pi$ rest frame. 

\item $\phi$ - the angle between the plane defined by the momenta of
  the $D^{(*)}$ and the $\pi$ and the plane defined by the momenta of
  the $\tau$ and $\bar{\nu}_\tau$ in the $\bar B$ rest frame.
\end{itemize}

In general, choices regarding the $\int d\Phi$ integral need to
balance two requirements. On the one hand, performing the analysis in
terms of several phase-space variables is experimentally
daunting. This is partly due to the complex modeling of correlated
background distributions, but also due to the difficulty of measuring
all the variables in the presence of unobservable neutrinos. On the
other hand, integration leads to cancellation of opposite-sign
contributions to the CP asymmetry in different regions of phase space.
Thus, our goal is to optimize the phase-space integration with
these considerations in mind.

First, we identify the integrals that make the asymmetry vanish. Since
the main source of strong phases in our method is interference between
excited charm mesons, integrals that reduce these interference terms
are undesirable.
In particular, since the phase of the Breit-Wigner amplitude varies as
a function of $m^2_{D^{(*)}\pi}$, the distribution of this variable is
critical for the analysis and must not be integrated over.
Experimentally, $m^2_{D^{(*)}\pi}$ is straightforward to evaluate,
since the 4-momenta of the $D^{(*)}$ and the $\pi$ are directly
measured. Furthermore, the $m^2_{D^{(*)}\pi}$ measurement resolution
is much smaller than the widths $\Gamma_{D^{**}_i}$ and the mass
differences $M_{D^{**}_i} - M_{D^{**}_j}$, which set the mass scale
over which the strong-phase difference varies significantly.

Next, we consider the angular variables.  A well-known fact in quantum
mechanics is that while the angular-momentum operator does not commute
with the momentum operator, $[\hat{\bf P}, \hat{\bf L}^2]\neq 0$, it
does commute with its square, $[\hat{\bf P}^2, \hat{\bf L^2}]= 0$.
Therefore, as long as we keep track of the directions of the $D^{**}$
daughter particles, we do not know the spin of the $D^{**}$,
allowing interference to take place.
On the other hand, in a gedanken experiment that cannot measure
momentum eigenstates but does measure the $\hat{\bf L}^2$ quantum number
of the $D^{(*)}\pi$ two-particle wave function,
interference between intermediate states of different spins is
forbidden by selection rules. Mathematically, this can be
understood from the orthogonality of the $Y_{lm}$ spherical harmonic
functions. Thus, we conclude that we must not integrate over the
entire $D^{(*)}\pi$ angular range defined by both $\theta_D$ and
$\phi$. In the general case, integration over one of the
phases does not completely cancel the asymmetry.

Since the strong phases come from the hadronic part
of the decay, integrating over the leptonic phase-space variables
$\theta_\tau$ and $q^2$ does not in principle cancel the CP
asymmetry. This is encouraging, since it is experimentally difficult
to measure $\theta_\tau$ and $\phi$. In practice, however, integration
over these angular variables does reduce the asymmetry. We study this
effect, as well as the extent to which it can be mitigated, in the
following section.
To summarize, we find that the experimentally simplest, nonvanishing
CP asymmetry is
\begin{align}\label{eq:Acp asymmetry}
\mathcal{A}_{CP}(m_{D^{(*)}\pi},\,\theta_D) 
\equiv \frac{\int d(\cos\theta_\tau )\, d\phi \, dq^2 \left(\left|\bar{\mathcal{A}}\right|^2-\left|\mathcal{A}\right|^2\right)}{\int d(\cos\theta_\tau) \, d\phi \, dq^2 \left(\left|\bar{\mathcal{A}}\right|^2+\left|\mathcal{A}\right|^2\right)}~.
\end{align}
In the case of $D^{**}\to D^*\pi$ decays, we assume that integration
over the decay angle of the $D^*\to D\pi$ or $D^*\to D\gamma$ decay is
performed, and hence implicitly sum over the $D^*$ helicity states.

\section{Toy model}\label{sec:Simplified Model}

We consider a toy model in order to illustrate
our method and obtain a rough estimation of the
asymmetry. In what follows we make the following assumptions:
\begin{enumerate}
\item
The solution to the $R(D^{(*)})$ puzzle originates from new degrees of
freedom that are heavier than the EW scale, and their effect can be
represented by non-renormalizable terms in the Lagrangian.
\item 
The NP modifies only $b\to c\tau^-\bar{\nu}_\tau$ transitions, while
$b\to c\ell^-\bar{\nu}_\ell$ is given by the SM.
\item
We neglect EW breaking effect in the NP operators, {\it i.e.}, we
assume that the NP terms are invariant under the $SU(3)_C\times
SU(2)_L\times U(1)_Y$ gauge symmetry of the SM.
In practice, this means that we ignore the vector operator $(\bar{\tau}_L \gamma^\mu \nu_{\tau_L})(\bar{c}_R \gamma_\mu b_{R})$.

\item
  There are no new states that are lighter than the weak scale. In
  particular, there are no light right-handed neutrinos.
\end{enumerate}

Under these assumptions, one finds that only four operators can break
lepton universality in $b\to c\ell^-\bar{\nu}_\ell$
transitions~\cite{Blanke:2017qan}:
\begin{align}\label{eq: Dim6 operators}
\mathcal{O}_{V_L^{(3)}} = (\bar{L}\gamma^\mu \tau^a
  L)(\bar{Q}\gamma_\mu \tau^a Q) ,\qquad &
\mathcal{O}_{S_R}=(\bar{e}L)(\bar{Q}d), \\
\mathcal{O}_{S_L}=(\bar{e}L)(\bar{u}Q) ,\qquad &
\mathcal{O}_{T}=(\bar{e}\sigma^{\mu\nu}L)(\bar{u}\sigma_{\mu\nu}Q).
\end{align}
Following standard notation, here $L$ and $e$ are the $SU(2)_L$
doublet and singlet lepton fields, and $Q,\,u$ and $d$ are the
$SU(2)_L$ doublet, up-singlet and down-singlet quark fields. Since the
SM operator is $\mathcal{O}_{SM} = \mathcal{O}_{V_{L^{(3)}}}$, the CP
asymmetry is not sensitive to a NP phase in the Wilson coefficient of
this operator. Therefore, we do not consider this operator further.
At the scale of the $B$-meson, it is more convenient to work in the
broken phase with a different linear combination of the remaining
operators. We use the following basis:
\begin{align}\label{eq:dim6operatorBroken}
\mathcal{O}_S = (\bar{\tau}_R\nu_{\tau_L})(\bar{c}b),\qquad
\mathcal{O}_P = (\bar{\tau}_R\nu_{\tau_L})(\bar{c}\gamma^5 b),\qquad
\mathcal{O}_T = (\bar{\tau}_R\sigma^{\mu\nu}\nu_{\tau_L})(\bar{c}\sigma_{\mu\nu}b).
\end{align}

We remark that our study is purely phenomenological, and we do not
attempt to address solutions to the $R(D^{(*)})$
anomaly. Nevertheless, it is known in the literature ({\it e.g.}
Ref.~\cite{Blanke:2017qan} and references therein) that if a single
mediator is responsible for the anomaly, then there are four possible
candidate mediators, labeled $W'_\mu \sim (1,3)_0,\,
U_\mu\sim(3,1)_{2/3},\,S\sim(3,1)_{-1/3},\,V_\mu\sim(3,2)_{-5/6}$. The main role of those mediators is to generate a significant contribution to $\mathcal{O}_{V^{(3)}_L}$, which tends to solve the $R(D^{(*)})$ anomaly by means of new physics. Except
for the case of $W'_\mu$, integrating out the mediator generically
leads to one or more of the operators
$\mathcal{O}_{S_R},\,\mathcal{O}_{S_L},\,\mathcal{O}_T$ being of the
same order of magnitude as the NP contribution to
$\mathcal{O}_{V^{(3)}_L}$.

In what follows we study the $D^*\pi\tau^-\bar{\nu}_\tau$ final state
of the $D^{**}$ decays. We discuss the $D\pi\tau^-\bar{\nu}_\tau$ final
state in Sec.~\ref{sec:conclusion}. For the purpose of this
proof-of-principle discussion, we make several simplifications.
While some have been mentioned above, we collect them all here for 
completeness:
\begin{enumerate}
	\item We assume that the observation of $D^*$ includes
          integration over the $D^*$ decay angle. Therefore, we
          neglect interference of different $D^*$ helicity states.

	\item
	We neglect CPV in the the SM.  This is
        exact up to tiny higher-order corrections to the tree-level SM
        process.

	\item We consider interference only between the narrow $D_1$ and
          $D_2^*$ resonances, ignoring the broad $D_1^*$, which also
          decays to $D^*\pi$. 
        As in the case of the ignored nonresonant amplitude discussed in
        Sec.~\ref{sec:CPintro}, the broad resonance contributes
        little over the small mass range covered by the narrow
        resonances.

	\item
	We use the Breit-Wigner approximation for the resonances, as
        shown explicitly in Eq.~\eqref{eq:Narrow width}. We expect
        this to be a good approximation close to the resonance peak,
        and become less precise farther from the peak. Corrections to
        this limit can be accommodated if needed. See details in, {\it
          e.g.}, the resonances section of~\cite{Patrignani:2016xqp} and references therein.

	\item We assume factorization of the hadronic current and
          the leptonic currents in Eqs.~(\ref{eq:B to D** tau nu
            amplitudes parametrized SM})--(\ref{eq:B to D** tau nu
            amplitudes parametrized NP}), $i.e.~\langle D^{** \, +}
          \tau^- {\bar \nu}_\tau | {\cal O} | {\bar B}^0 \rangle
          \simeq \langle D^{** \, +} | {\cal O}_{q} | {\bar B}^0
          \rangle \langle \tau^- {\bar \nu}_\tau | {\cal O}_{\ell}
          |0\rangle$.
\item
We calculate the leptonic currents to leading order in perturbation theory.
\item We calculate the $\bar{B}\to D^{**}$ transition 
        to leading order in the heavy quark limit, namely,
          neglecting corrections of order $\Lambda_{\rm
            QCD}/m_c$. This assumption has two implications. First, as
          discussed above, we set the non-Breit-Wigner strong phases
          to zero. Second, we set all form factors to be the same and
          equal to a single Isgur-Wise function. The hadronic matrix
          elements $\braket{D^{**}_i | \mathcal{O}| \bar{B}}$ are
          given explicitly in App.~\ref{app:B to D** HQET}. Subleading
          $1/m_Q$ and $\alpha_s$ corrections are given
          in~\cite{Bernlochner:2017jxt}.

	\item For the $D^{**}$ decay amplitude we use an approximate model
	inspired by leading-order heavy quark effective theory (HQET). Details are given in App.~\ref{app:D** to D*pi decay}.
\end{enumerate}

It is important to note that these simplifications do not change the
major conclusions of our study. Furthermore, as discussed in
Sec.~\ref{sec:RealExperiment}, they pose no limitation for actual
analysis of experimental data.

\subsection{Results and cross checks}
\label{sec:results-checks}

Using the above simplified model, we calculate the CP asymmetry of
Eq.~\eqref{eq:Acp asymmetry}. For this purpose, we set the Wilson
coefficient of one of the operators in
Eq.~\eqref{eq:dim6operatorBroken} to $C^{NP}_A = 0.15 (1+i)\,C_{SM}$
(where $A=S/P/T)$, while setting the others to zero. The choice of
this value is arbitrary, but motivated by the $\sim 30\%$ enhancement
of the central values of $R(D^{(*)})$ with respect to the SM
expectation.  We choose $(1+i)$ to obtain an arbitrary order-one 
value for $\varphi^{\rm NP}$.

It is unrealistic that the UV physics that solves the $R(D^{(*)})$
anomaly generates just a single operator as is assumed in our
phenomenological study.  
However, we explicitly checked that using a generic
linear combination of those operators, particularly either one of
the motivated combinations $\mathcal{O}_{S_R}$ or $\mathcal{O}_{S_L}$,
does not significantly modify our results.

\begin{figure}[t]%
	\centering
	\subfloat[][]{\includegraphics[scale=0.66]{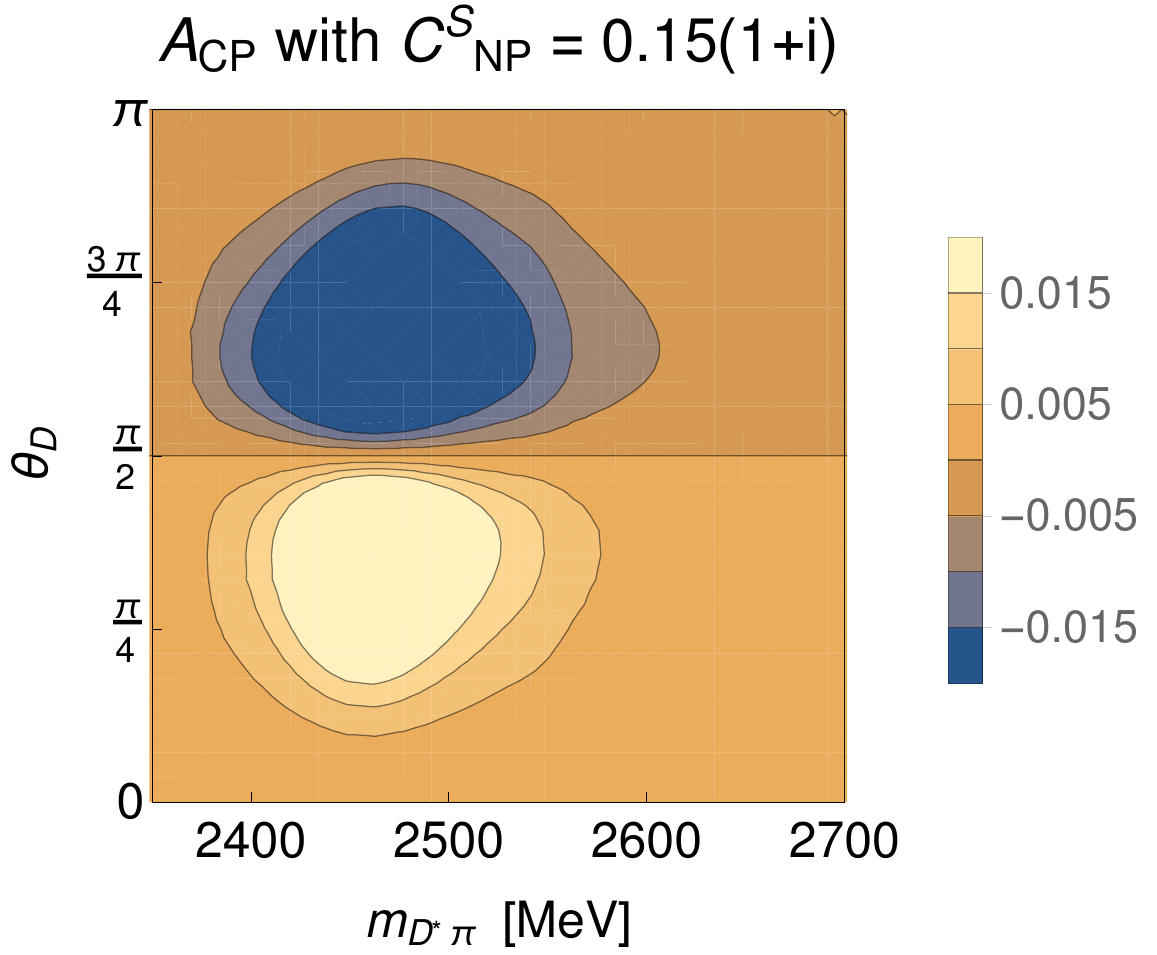}\label{fig:Money_S}}%
	\qquad
	\subfloat[][]{\includegraphics[scale=0.66]{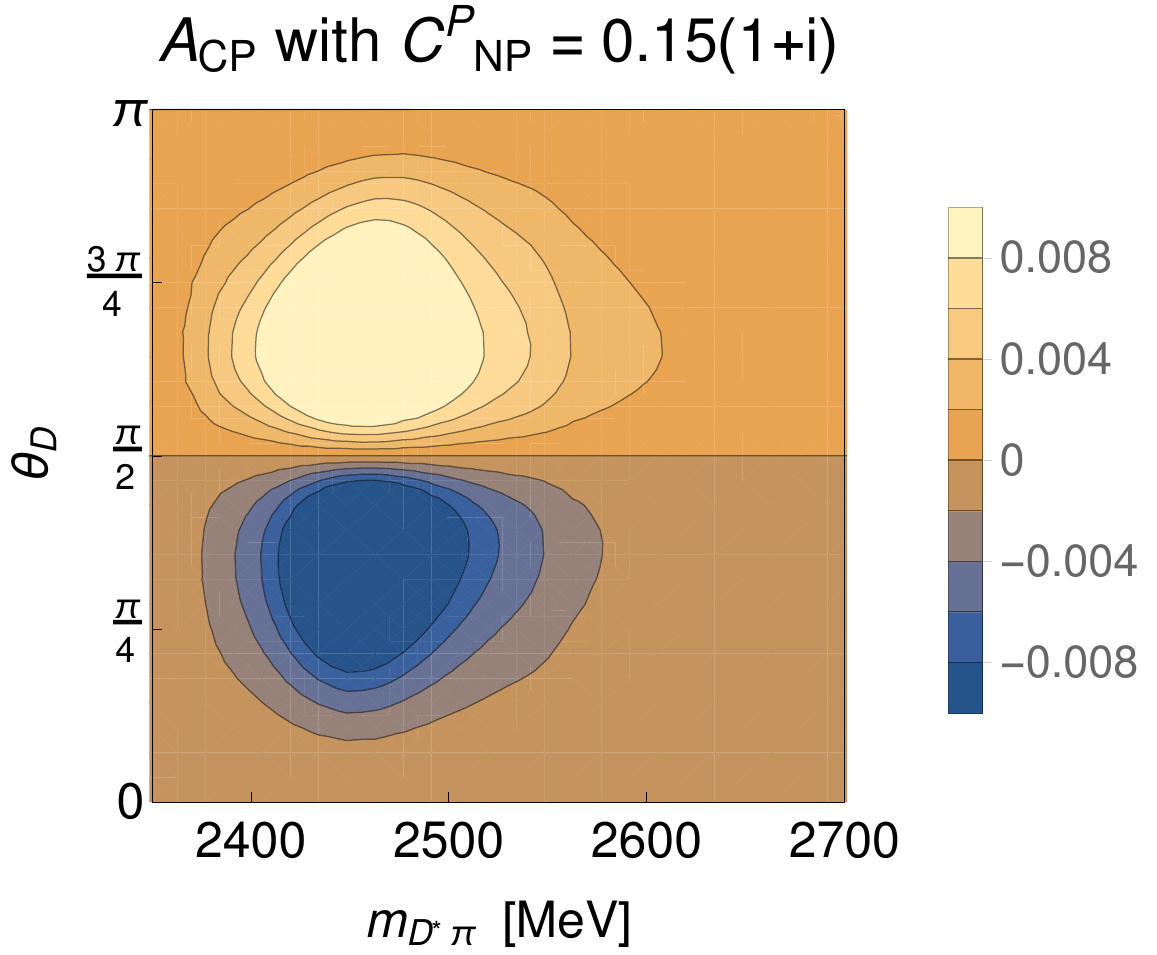}\label{fig:Money_PS}}%
	\qquad
	\subfloat[][]{\includegraphics[scale=0.66]{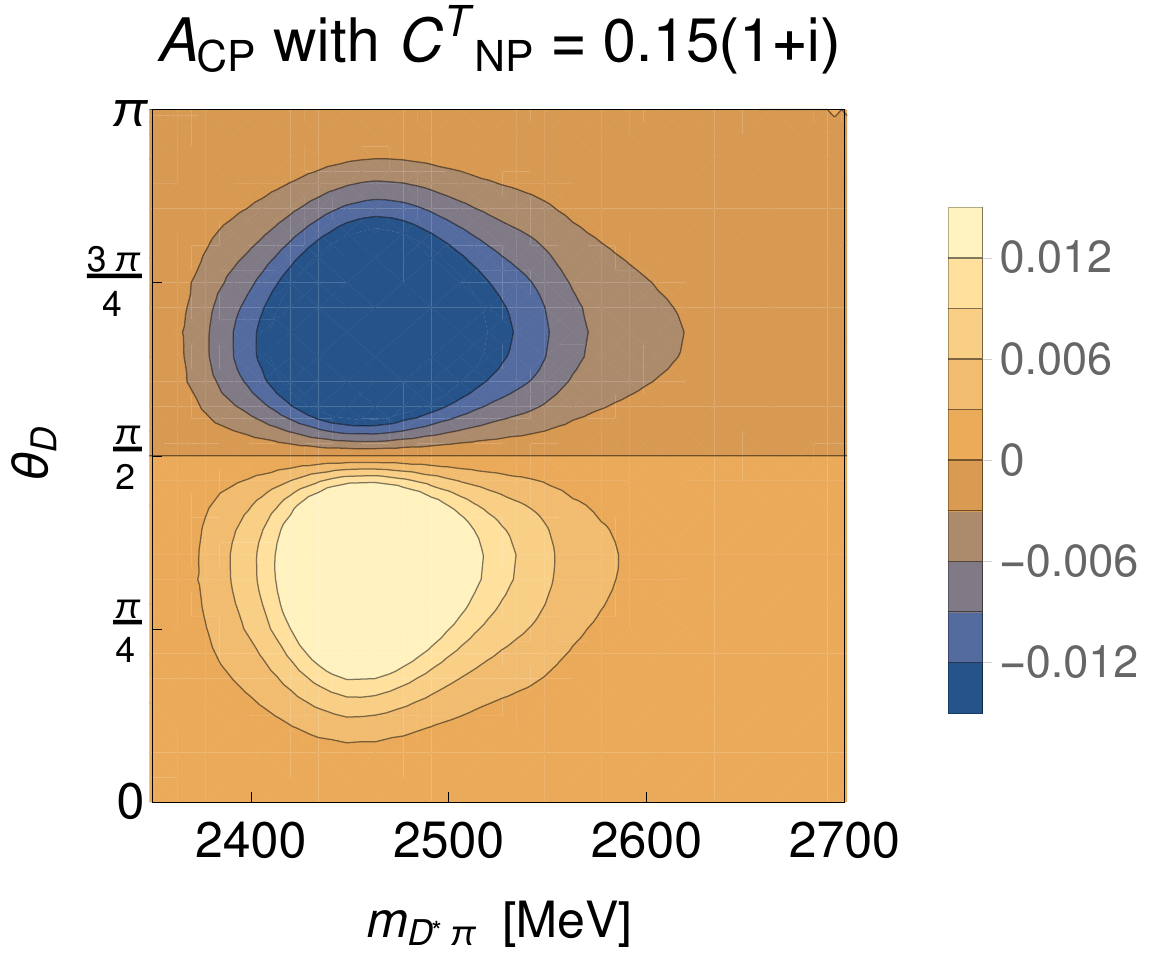}\label{fig:Money_T}}%
	\caption{The CP asymmetry of Eq.~\eqref{eq:Acp asymmetry} as a
          function of the $D^{*}\pi$ invariant mass $m_{D^{(*)}\pi}$
          and the $D^{**}$ decay angle $\theta_D$ for a (a) scalar,
          (b) pseudoscalar, and (c) tensor NP operator.}
	\label{fig:MoneyPlot1}%
\end{figure}

In Fig.~\ref{fig:MoneyPlot1} we plot the asymmetry of
Eq.~\eqref{eq:Acp asymmetry} as a function of the $D^{*}\pi$ invariant
mass $m_{D^{(*)}\pi}$ and the $D^{**}$ decay angle $\theta_D$ for the
three NP operators. We find the asymmetry to be of order one
percent.

\begin{figure}[t]%
	\centering
	\subfloat[][]{\includegraphics[scale=0.68]{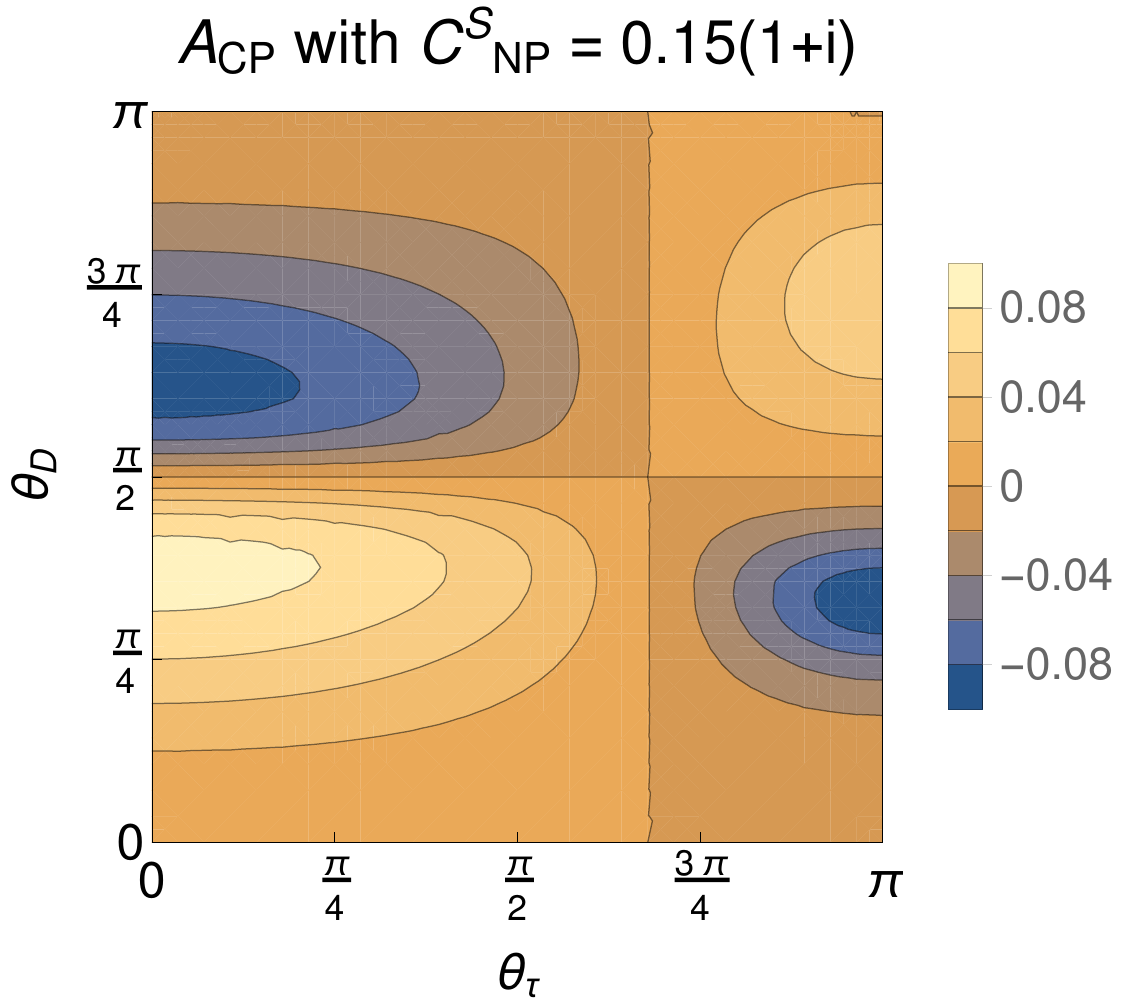}\label{fig:Fig2a}}%
	\qquad
	\subfloat[][]{\includegraphics[scale=0.68]{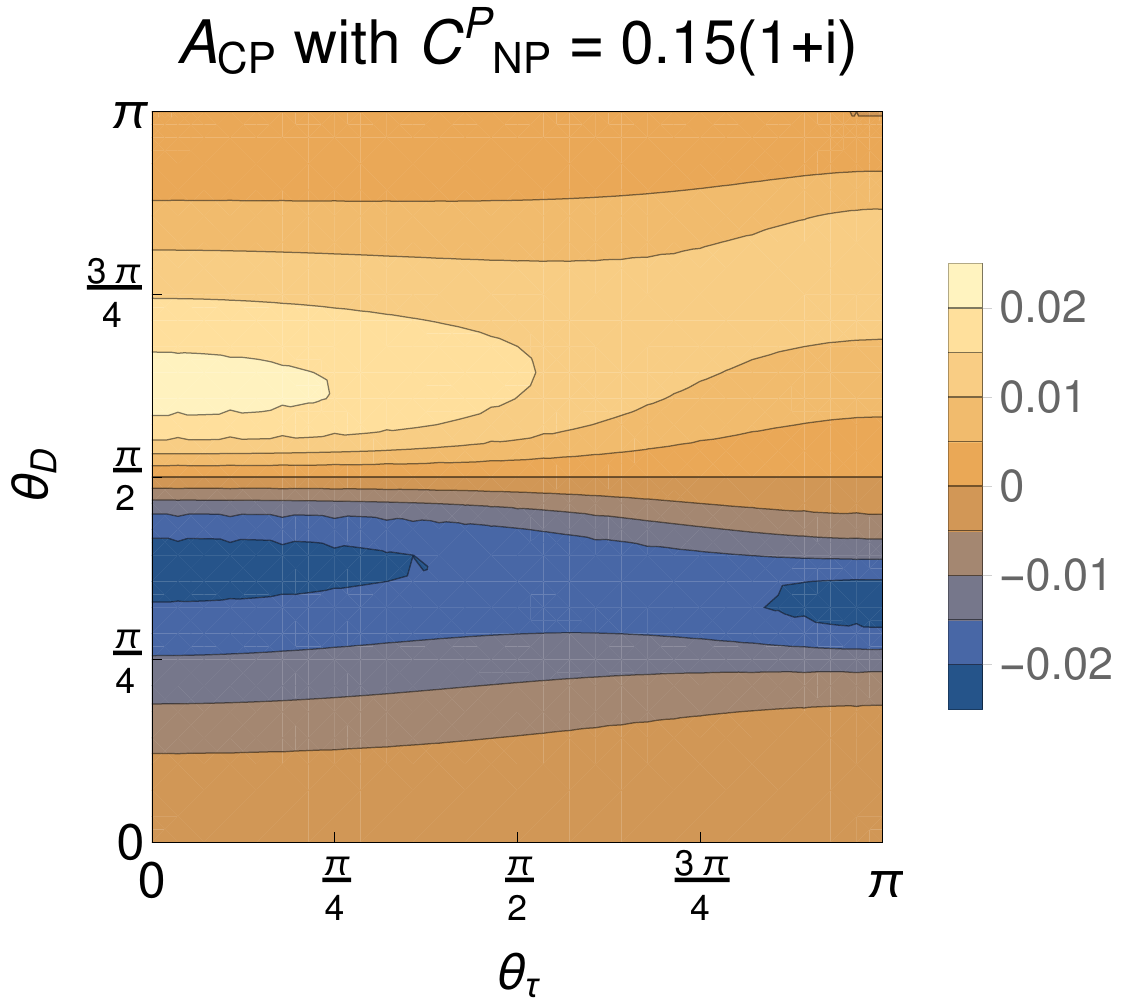}\label{fig:Fig2b}}%
	\qquad
	\subfloat[][]{\includegraphics[scale=0.68]{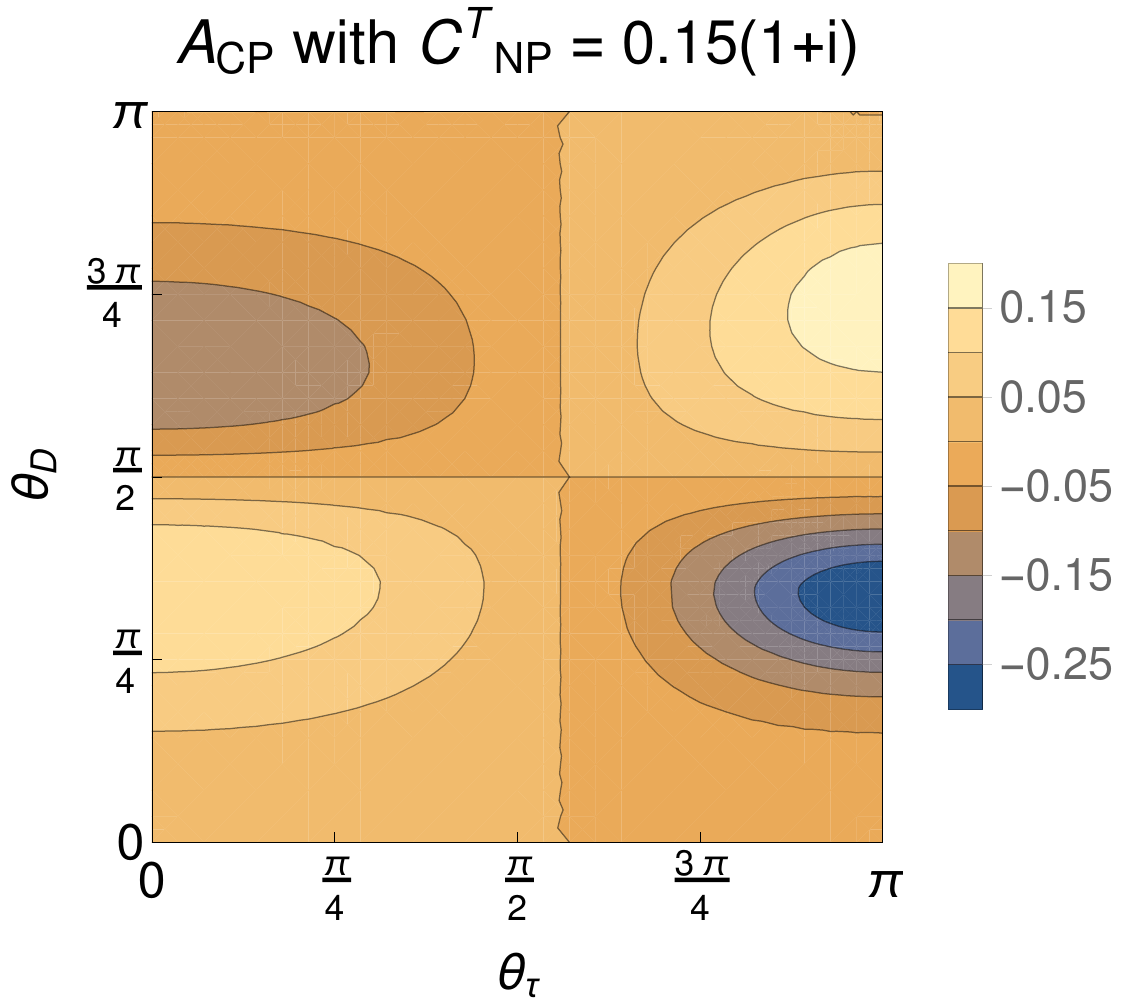}\label{fig:Fig2c}}%
	\caption{The CP asymmetry of Eq.~\eqref{eq:Acp asymmetry} as a
          function of the angles $\theta_D$ and $\theta_\tau$ for a
          fixed value of the $D^{*}\pi$ invariant mass, for a (a)
          scalar, (b) pseudoscalar, and (c) tensor NP operator.
  }
	\label{fig:ThetaL Dalitz plot}%
\end{figure}

\begin{figure}[t]%
	\centering
	\subfloat[][]{\includegraphics[scale=0.68]{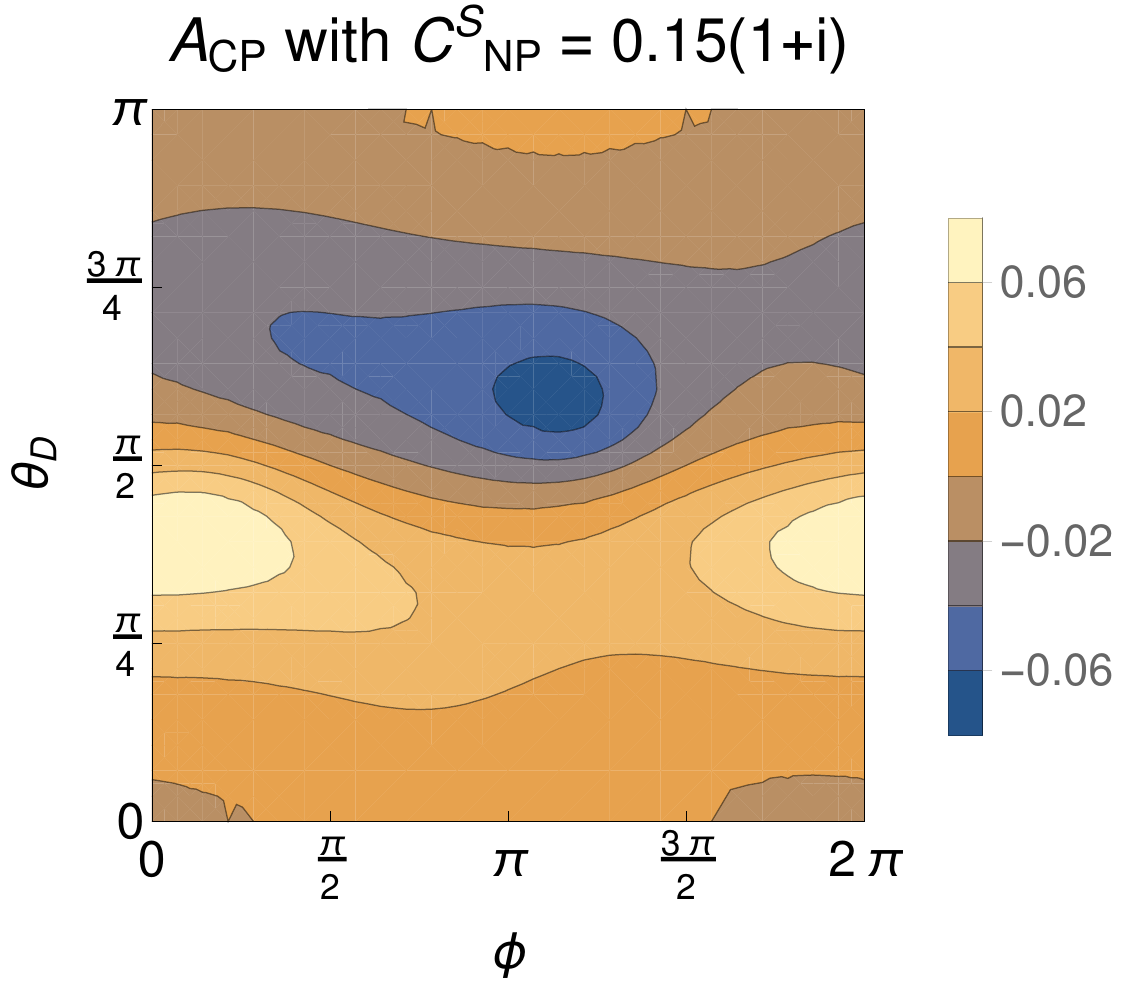}\label{fig:Fig3a}}%
	\qquad
	\subfloat[][]{\includegraphics[scale=0.68]{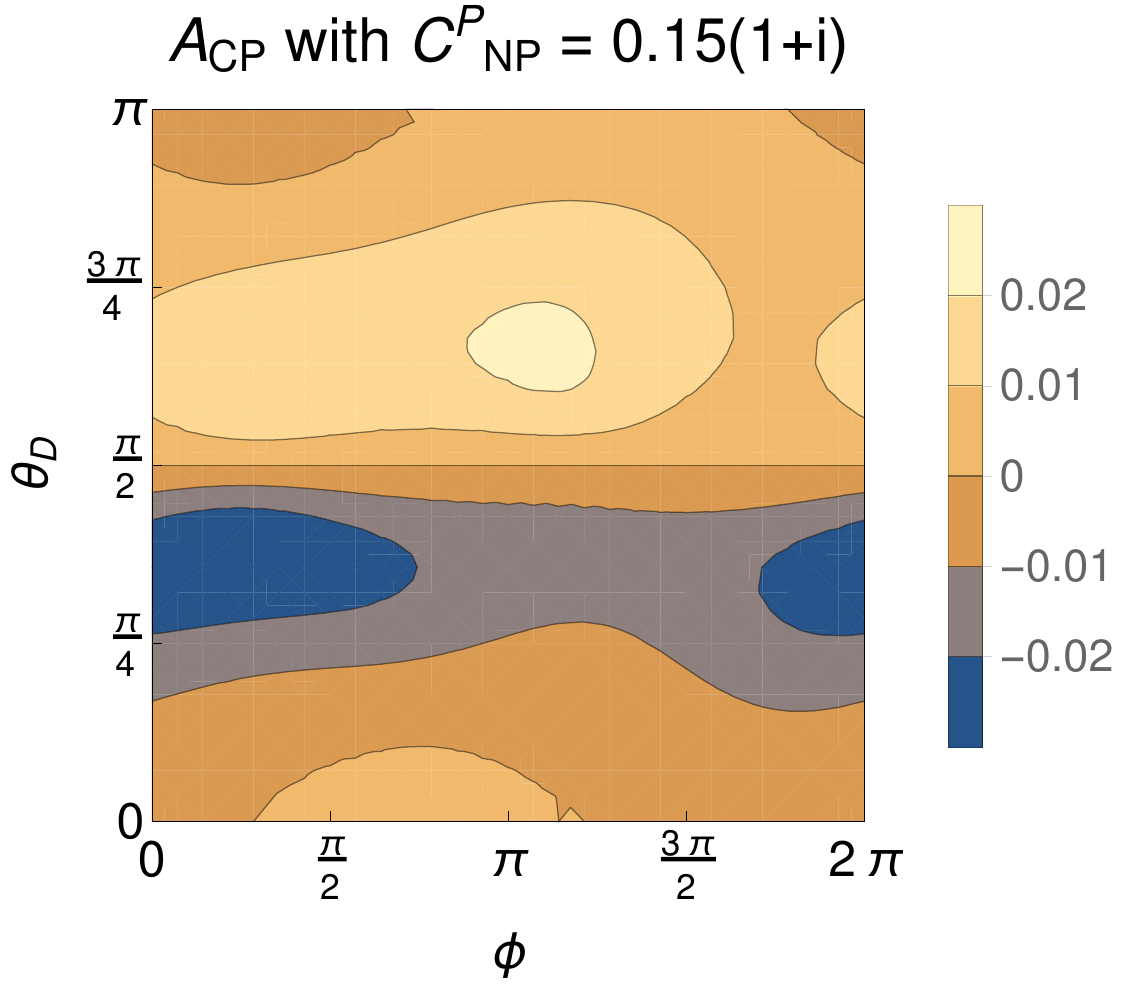}\label{fig:Fig3b}}%
	\qquad
	\subfloat[][]{\includegraphics[scale=0.68]{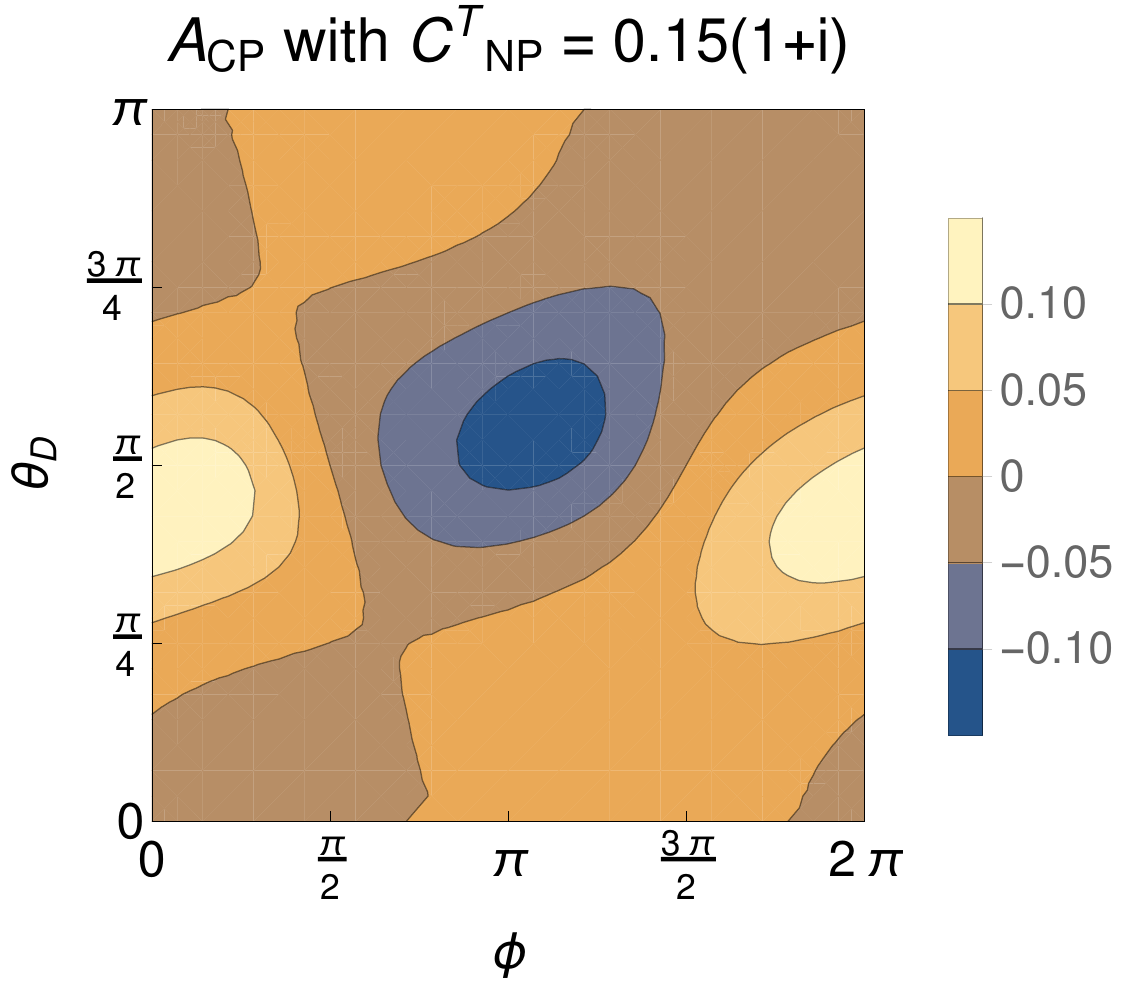}\label{fig:Fig3c}}%
	\caption{The CP asymmetry of Eq.~\eqref{eq:Acp asymmetry} as a
		function of the angles $\theta_D$ and $\phi$ for a
		fixed value of the $D^{*}\pi$ invariant mass, for a (a)
		scalar, (b) pseudoscalar, and (c) tensor NP operator.
	\label{fig:phi Dalitz plot}%
}
\end{figure}

In addition, we study the implication of having partial knowledge of
the $\tau$ angular distribution. To simplify this study, we take a
representative value of the $D^*\pi$ invariant mass, fixing it to be
between the peaks of the two resonances, $m_{D^*\pi} = (M(D_1)
+M(D_2^*))/2$, where the Breit-Wigner phase difference is large. 
We then plot the asymmetry in the plane of $\theta_D$
vs. either $\theta_\tau$ (Fig.~\ref{fig:ThetaL Dalitz plot}) or $\phi$
(Fig.~\ref{fig:phi Dalitz plot}), after integrating over the remaining
variables.
As shown in these figures, retaining the $\theta_\tau$ or $\phi$ dependence
leads to up to an order of magnitude enhancement in the asymmetry.
As mentioned in the introduction, one objective of our study is to
propose a CP-violation analysis which, in contrast to previous
proposals, does not require a full angular analysis. Nonetheless, one
can slightly relax this requirement and observe a larger asymmetry
when measuring two of the three phase-space angles. We discuss the 
experimental aspects of this approach in Sec.~\ref{sec:experimental}.


\begin{figure}[t]%
	\centering
	\subfloat[][]{\includegraphics[scale=0.63]{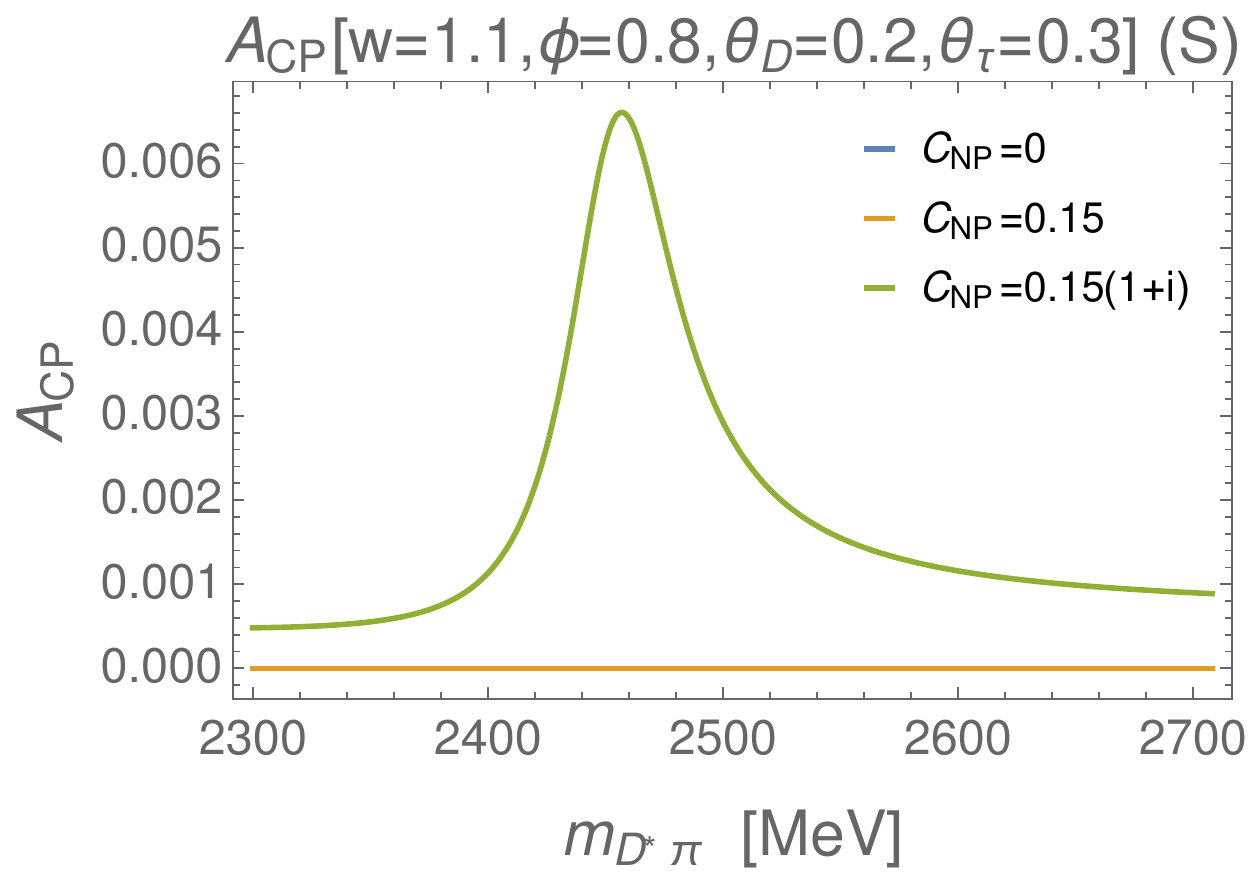}\label{fig:Fig4a}}%
	\quad
	\subfloat[][]{\includegraphics[scale=0.63]{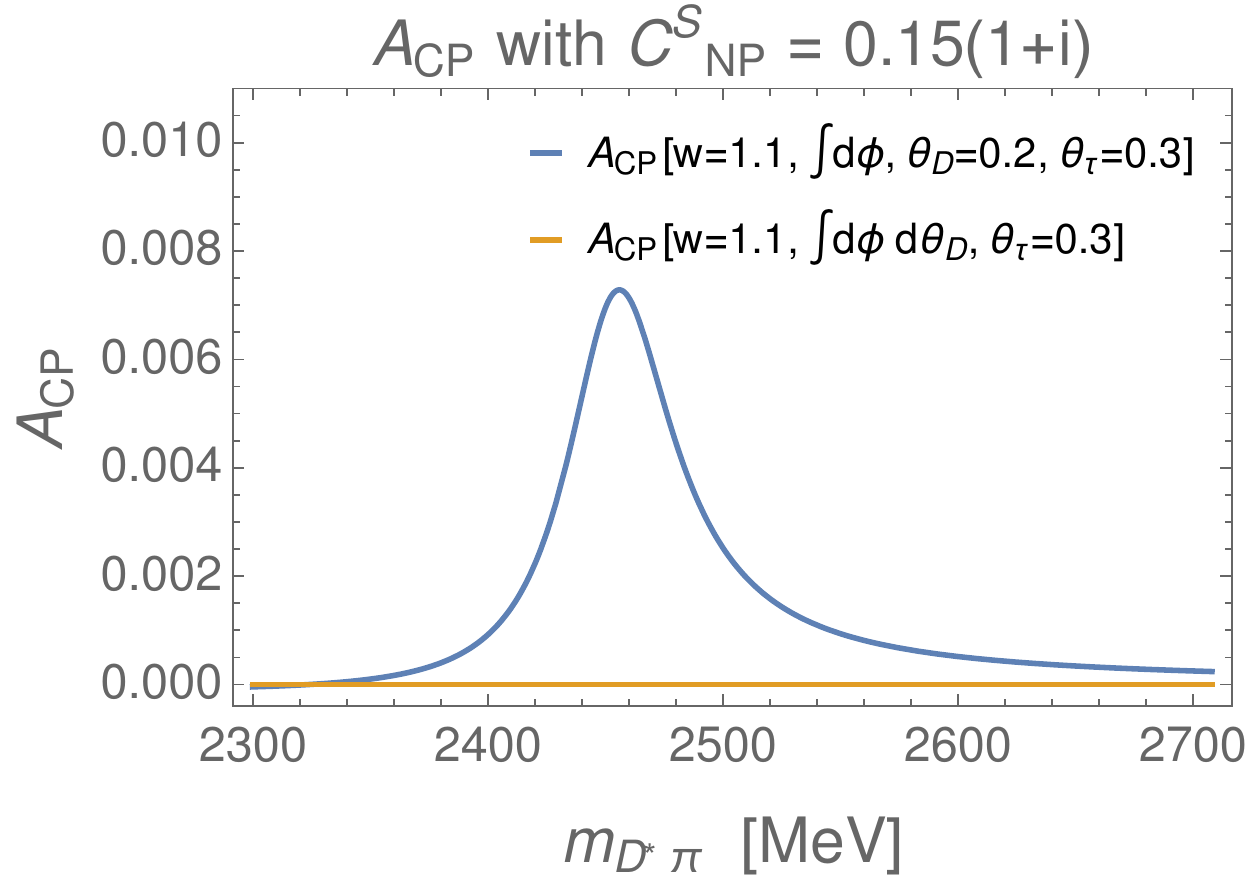}\label{fig:Fig4b}}%
	\caption{Representative cross checks, showing the CP-asymmetry
          in an arbitrary point in phase spcae for the case of a
          scalar NP operator. (a) Results obtained with (blue) no NP,
          (orange) NP without CP violation, and (green) NP with CP
          violation. (b) Results obtained with NP with CP violation,
          after integration over (blue) $\phi$ or (orange) $\phi$ and
          $\theta_D$.
	}
	\label{fig:validity}%
\end{figure}

In order to verify the validity of our numerical results, we
performed a set of cross checks. For a random point in phase-space, we
verified that the CP asymmetry vanishes when the NP Wilson
coefficients are set to be real, {\it i.e.}, $\varphi^{\rm
  NP}=0$. This is shown in Fig.~\ref{fig:Fig4a} for the scalar
operator, and similar results are obtained for the pseudoscalar and
tensor cases.
For complex Wilson coefficients, we verified that before phase-space
integration there is a $\phi$-dependent CP asymmetry even in the case
of a single mediator, as in Ref.~\cite{Duraisamy:2013kcw}. A remnant
of this effect is seen in the off-resonance sidebands of the green
curve in Fig.~\ref{fig:Fig4a}.
As can be seen in Fig.~\ref{fig:Fig4b}, this asymmetry vanishes after
integrating over $\phi$ (blue curve). Finally, as discussed above, we
have verified that for a fixed arbitrary angle $\theta_\tau$, the asymmetry
vanishes after integration over the hadronic angles $\theta_D$ and
$\phi$. This is shown by the orange curve in Fig.~\ref{fig:Fig4b}.

\section{A detailed extension to real life experiment}
\label{sec:RealExperiment}

In the previous section we listed the assumptions used in our
proof-of-principle study. In an actual analysis of collider data,
obtaining precise measurement of the asymmetry requires replacing
these assumptions with experimental results with properly evaluated
uncertainties. In Sec.~\ref{sec:theory-cons} we discuss the validity
of these assumptions and explain that they do not change the nature of
our conclusions, particularly the general magnitude of the asymmetry
for given values of the complex NP Wilson coefficients.
Experimental considerations are discussed in Sec.~\ref{sec:experimental}.

\subsection{Theoretical considerations}
\label{sec:theory-cons}

\begin{enumerate}
	        \item As long as we integrate over the $D^*\to D\pi$
                  angular distribution, there is no interference
                  between different $D^*$ helicity states. Therefore,
                  the implicit incoherent sum over different
                  helicities in Eq.~\eqref{eq:Acp asymmetry} is
                  precise. Nevertheless, retaining some information about the $D\pi$ angular distribution would generally give
                  rise to interference between different helicities of
                  the $D^*$. As $D_2^*$ decays only to transverse
                  $D^*$ (see App.~\ref{app:D** to D*pi decay}), this
                  would result in a somewhat larger asymmetry.
                  Another effect related to the $D^*$ angular
                  distribution that might enhance the asymmetry
                  somewhat, and which we do not study here, is similar
                  to that of the $D$ angular distribution studied
                  in Ref.~\cite{Duraisamy:2013kcw}.

	        \item The assumption of tiny CPV within the SM is
                  straightforward, and does not require additional discussion.
%
	        
	\item In our calculations of the asymmetry we considered only
          the $D_1$ and $D_2^*$ resonances.  The only additional
          resonance that can contribute to the $D^*\pi$ final state is
          $D_1^*$. Since this state is very broad, its contribution to
          the observed final state across the narrow $m_{D^{(*)}\pi}$
          range defined by the $D_1$ and $D_2^*$ is small, and its
          phase varies only little. As a result, it does not affect
          our study significantly. It is advisable, however, to
          account for the $D_1^*$, along with nonresonant and
          background contributions, in the experimental data
          analysis. The experimental analysis would anyway obtain the
          parameters of all the $D^{**}$ resonances and the relevant
          form factors from the control studies described in
          Sec.~\ref{sec:CPintro}, particularly with $\bar{B}\to
          D^{**}\ell^- \bar{\nu}_\ell$~\cite{Aubert:2008ea}. In
          addition, matrix elements for the $\bar{B}\to D_1^*$
          transition can be taken from~\cite{Bernlochner:2017jxt}.

          \item Although we use the Breit-Wigner approximation, the
            shape of the resonance, and in particular the $M_{D^*\pi}$
            dependence of the strong phase, should be measured as one of
            the control studies.
          
          \item To leading order in HQET, all strong phases
            $\delta_i^{SM},\,\delta_i^{NP}$ are equal. Moreover,
            strong phases are expected to be small when the final
            state has only one hadron, as in this case, due to the
            absence of rescattering. In any case, the phase-space
            dependence of these phases is small, and thus does not
            lead to cancelation of the asymmetry.  Finally, we note
            that these phases can be measured in $\bar{B}\to
            D^{**}\ell^-\bar{\nu}_\ell$. Therefore, deviations from
            the assumptions outlined here do not change the
            conclusions of our work.

          \item Corrections to the factorization assumption and NLO
            corrections to the leptonic currents are $\alpha_{EM}$
            suppressed and thus negligible.
          
		\item NLO corrections to $\bar{B}\to D^{**}$
          transition form factors are  as large as tens of percent. Therefore $\mathcal{O}(1)$ corrections to the expected asymmetry arising from these terms are expected.  These corrections can be taken from
        Ref.~\cite{Bernlochner:2017jxt} and studied in $\bar{B}\to D^{**}\ell^-\bar{\nu}_\ell$ decays. In any case, they are not expected to significantly change the global picture that arises from our
        study. For instance, as pointed out in~\cite{Bernlochner:2017jxt} LO fails to predict the ratio $BR(\bar{B}\to D_2^* \ell \bar{\nu})/BR(\bar{B}\to D_1 \ell \bar{\nu})$. We checked explicitly that correcting for this discrepancy modifies our main results by $\mathcal{O}(10\%)$. 

        \item We use LO HQET for modeling the $D^{**}$ decay. Large
          corrections to this approximation are expected in purely
          charmed systems. 
		In particular, it is
                predicted~\cite{Lu:1991px,Falk:1992cx} that $D_1 -
                D_1^*$ mixing leads to significant S-wave
                contribution. As can be seen from Eq.~\eqref{eqApp:D1
                  amplitudes}, this leads to enhancement of the
                helicity-amplitude ratio
                $|\mathcal{A}^1_{10}|/|\mathcal{A}^1_{00}|$ (see
                App.~\ref{app:D** to D*pi decay}).
                Since different helicity amplitudes do not interfere, and
                $\mathcal{A}^2_{00}=0$ due to selection rules, a
                larger value of
                $|\mathcal{A}^1_{10}|/|\mathcal{A}^1_{00}|$ enhances
                the interference of $D_1$ and $D_2^*$, increasing the
                asymmetry. Thus, our use of LO HQET leads to a
                conservative estimate of the CP-violating signal.
	          In any case, the modeling of the $D^{**}$ decay will be
	          improved by precise measurements of the helicity amplitudes at
	          Belle~II and LHCb, as part of the control studies.

\end{enumerate}

\subsection{Experimental considerations}
\label{sec:experimental}

\begin{enumerate}
\item First, it is desirable to estimate the achievable
uncertainty on the asymmetry. Such an estimate is bound to be highly
inaccurate, due to lack of any experimental studies of $\bar B\to
D^{**}\tau^-\bar{\nu}_\tau$. Nonetheless, one can gain some insignt
from a study of $\bar B\to D^{**}\ell^-\bar{\nu}_\ell$ by
BABAR~\cite{Aubert:2008ea}, performed with full hadronic
reconstruction of the other $B$ meson in the event. 

For example, BABAR find $165\pm 18$ events in the channel $B^-\to
D_1^0\ell^-\bar{\nu}_\ell$.
At Belle~II, the integrated luminosity will be about 100 times
larger. However, the branching fractions for $\bar B\to
D^{**}\tau^-\bar{\nu}_\tau$ are expected to be about 10 times smaller
than those of $\bar B\to
D^{**}\ell^-\bar{\nu}_\ell$~\cite{Bernlochner:2016bci}. Hence, a
naive scaling of the BABAR result to Belle~II yields a $B^-\to
D_1^0\ell^-\bar{\nu}_\ell$ signal of $(165 \times 10) \pm (18\times \sqrt{10})$
events. This assumes that the signal efficiency and
signal-to-background ratio remain as in
Ref.~\cite{Aubert:2008ea}. There is no reason to think that these assumptions are correct, since the
two detectors, integrated luminosities, and analysis optimization
procedures are very different. 
The different signal-to-background ratios can be approximately
corrected for. For this purpose, we note that Fig.~(1a) of
Ref.~\cite{Aubert:2008ea} indicates a background yield of about 30
events under the $D_1$ and $D_2^*$ peaks. If we naively assume that
this background has negligible impact on the signal yield uncertainty
in Ref.~\cite{Aubert:2008ea} and that it will become 3000 events in
the Belle~II analysis due to the 100-fold increase in integrated
luminosity, then the expected uncertainty on the $B^-\to
D_1^0\tau^-\bar{\nu}_\tau$ signal yield at Belle~II becomes about
$\sqrt{18^2 \times 10 + 3000}\approx 80$~events. From this, one finds
that the uncertainty on a phase-space-integrated asymmetry would be
5\%.

While keeing in mind the caveats about the large inaccuracy of this
uncertainty estimate, we note that full exploitation of Belle~II data
would include also $\bar B^0$ decays, additional $D^{**}$ resonances,
and additional methods for reconstruction of the other $B$ meson in
the event, reducing the overall uncertianty. Furthermore, our estimate
pertains only to Belle~II, while LHCb is also likely to contribute
significantly to this measurement.

\item In Sec.~\ref{sec:results-checks} we showed that analyzing the CP
  asymmetry in terms of $\theta_\tau$, in addition to $\theta_D$,
  helps avoid cancelations and results in a large increase of the
  asymmetry.  Thus, it is important to understand whether
  $\theta_\tau$ can be determined with sufficient precision despite
  the unobservable neutrinos. We show here that this can be done at
  Belle~II, using the momentum $\vec p_\ell$ of the observed light
  lepton produced in the leptonic decay $\tau^-\to \ell^-\nu_\tau
  \bar\nu_\ell$. The kinematic constraints of the $e^+e^-\to B\bar B$
  production process provide information about the 3-momentum $\vec
  p_B$ of the $\bar B$ meson is known. As a result, the 3-momentum
  $\vec q$ of the $\tau^-\bar\nu$ system is determined to within about
  30~MeV or 300~MeV, depending on whether the other $B$ meson in the
  event is fully reconstructed via a hadronic
  decay~\cite{Lees:2012xj,Lees:2013uzd,Huschle:2015rga,Hirose:2016wfn}
  or partially reconstructed in a semileptonic
  decay~\cite{Sato:2016svk}. LHCb has also demonstrated the ability to
  measure $\vec q$ with some
  precision~\cite{Aaij:2015yra,Aaij:2017uff}. Knowledge of $\vec
  p_\ell$ and $\vec q$ enables measurement of $\theta_\ell$, the angle
  between $\vec p_\ell$ and $-\vec p_B$ in the $\tau^-\bar\nu$ rest
  frame.

By simulating the kinematics of the full decay chain using
EvtGen~\cite{Lange:2001uf} within the Belle~II software framework, we
find that knowledge of $\theta_\ell$ gives $\theta_\tau$ to within an
uncertainty of about $\pi/4$. As can be seen from Fig.~\ref{fig:ThetaL
  Dalitz plot}, this precision is sufficient for significantly
reducing the cancelation that would otherwise arise from integration
over $\theta_\tau$. Thus, observation of large asymmetry values, close
to those seen in Fig.~\ref{fig:ThetaL Dalitz plot} for the scalar and
tensor cases, is in principle possible.

In the semihadronic decays $\tau^-\to \pi^-(n\pi^0)\nu_\tau$ and
$\tau^-\to \pi^-\pi^+\pi^-(n\pi^0)\nu_\tau$ (where $(n\pi^0)$ stands
for additional neutral pions that may or may not be reconstructed),
the corresponding angles $\theta_\pi$ and $\theta_{3\pi}$,
respectively, should give a somewhat more precise estimate of
$\theta_\tau$ than that obtained from $\theta_\ell$, thanks to the
presence of only one neutrino in the decay. In $\tau^-\to
\pi^-\pi^+\pi^-(n\pi^0)\nu_\tau$, vertexing provides additional
information on $\theta_\tau$. At Belle~II, the $\tau$ flies an average
distance of $50~\mu$m before decaying, while the position resolutions
on the $B$ decay vertex and on the $\pi^-\pi^+\pi^-$ production vertex
are about $25~\mu$m each~\cite{Abe:2010gxa,Adamczyk:2016tta}. This too
enables an estimate of $\theta_\tau$ with an uncertainty of order
$\pi/4$.

The locations of the $B$ and $\tau$ decay vertices also
allow an estimate of the angle $\phi$ with similar precision. We note
that LHCb has already demonstrated successful use of the $\tau^-\to
\pi^-\pi^+\pi^-(n\pi^0)\nu_\tau$ decay vertex to study $\bar B\to
D^{(*)}\tau^-\bar{\nu}_\tau$~\cite{Aaij:2017uff}. Thus, LHCb may be
able to determine $\phi$ and $\theta_\tau$ in this way with better
precision than Belle~II.

\item 
	Finally, we note that as for any
  multibody final state, one has to account for the dependence of the
  reconstruction efficiency on the phase-space variables. This applies
  to both the measured and the integrated variables.

\end{enumerate}

\section{Summary and conclusions}\label{sec:conclusion}
In this paper we suggest a new method to study CP violation in
$\bar{B}\to D^{**}(\to D^{(*)}\pi)\tau^- \bar{\nu}_\tau$ decays. Our
motivation is based on the so-called $R(D^{(*)})$ anomaly of lepton
flavor non-universality in $\bar{B}\to D^{(*)}\tau^- \bar{\nu}_\tau$
decays. If this anomaly is indeed a result of physics beyond the SM,
it is natural to assume that the new physics amplitude may also have an
order-one CP violating phase with respect to the SM amplitude.

The source of strong-phase difference in our scheme is interference
between intermediate $D^{**}$ resonances. In the Breit-Wigner
approximation, it is transparent that this phase difference obtains
large values when the invariant mass of the $D^{**}$ final state is in
the range between the interfering resonance peaks.

We study in detail the case of interference between the narrow
resonances $D_1$ and $D_2^*$, which decay to the common final state
$D^*\pi$. While the $D_1^*$ also decays to $D^*\pi$, we do not expect
it to contribute significantly to the asymmetry due to its large
width, which results in small overlap with the narrow resonances.
The $D\pi$ final state may also be used for this measurement.  In this
case, interference takes place only between the $D_2^*$ and the
$D_0^*$. The large width of the $D_0^*$ again leads to a small
expected asymmetry.

We find that an order-one CP-violating phase in the new-physics
amplitude results in an order-percent CP asymmetry (Eq.~\eqref{eq:Acp
  asymmetry}) when integrating over the $\tau^-\bar{\nu}_\tau$
kinematics. We also observe that partial measurement of the direction
of the $\tau^-$ momentum leads to an order-of-magnitude enhancement of the
asymmetry. We outline how such a measurement can be performed at
Belle~II and LHCb.
Our main results are summarized in Figs.~\ref{fig:MoneyPlot1},
\ref{fig:ThetaL Dalitz plot}, and~\ref{fig:phi Dalitz plot}.

We make several approximations that help clarify the physical effect
and its kinematical dependence while also giving the correct order of
magnitude for the asymmetry.  We discuss the use of control studies,
particularly with $\bar B\to D^{**}\ell^-\bar{\nu}_\ell$ and $\bar B\to
D^{**} \pi^-$, for the purpose of studying these approximations,
improving upon them, and obtaining the related systematic uncertainties.

Our main goal is to study an observable that can probe NP by observing
a nonvanishing CP asymmetry. However, we note that a full analysis,
which includes measurement of $R(D^{**})$, the strong phases, and form
factors, yields the value of the underlying CPV phase $\varphi^{\rm
  NP}$. Clearly, initial observation of an asymmetry would provide the
motivation for such an analysis.

We discuss the uncertainty with which the proposed CP asymmetry can be
measured at Belle~II, using a BABAR study of $B^-\to
D_1^0\ell^-\bar{\nu}_\ell$ with hadronic reconstruction of the other
$B$ meson in the event. We find that for this channel alone, the
uncertainty is about 5\%. Nonetheless, we caution that such an
estimate is highly inaccurate, and encourage more detailed
experimental studies to obtain a better estimate, at both Belle~II and
LHCb.

As a last remark, we note that while current measurements motivate
searching for new physics and CP violation in $\bar B\to
D^{**}\tau^-\bar{\nu}_\tau$, our method can also be applied to the
search for a CP asymmetry in any other $\bar B$ decay involving
$D^{**}$ mesons, including $\bar B\to D^{**}\ell^-\bar{\nu}_\ell$.  In
this case, one loses the benefit of using $\bar B\to
D^{**}\ell^-\bar{\nu}_\ell$ for control studies. This is likely to
result in reduced sensitivity, but cannot create a fake CP
asymmetry.\footnote{As in any measurement of a CP-odd observable,
  systematic effects related to the CP asymmetry of the detector need
  to be accounted for.} Such a search can provide a powerful test for
CP violation in semileptonic $\bar B$ decays.

\acknowledgments We thank Diptimoy Ghosh, Yosef Nir and Zoltan Ligeti for useful
discussions. This research is supported in part by the United
States-Israel Binational Science Foundation (BSF), Jerusalem, Israel
(grant numbers 2014230 and 2016113). YG is supported in part by the
NSF grant PHY-1316222.

\appendix

\section{$B\to D^{**}$ HQET}\label{app:B to D** HQET}
For this paper to be self contained, we give the leading order HQET  terms for the $B\to D^{**}$ matrix elements. For the $D_1$ meson we have
\begin{align}
\frac{\braket{D_1(v',\epsilon)|S |B(v) }}{\sqrt{M_D M_B}} 
& = -\sqrt{\frac{2}{3}} \tau(w) (1+w)\epsilon^*\cdot v  ~,\\
\frac{\braket{D_1(v',\epsilon)|P |B(v) }}{\sqrt{M_D M_B}} 
& = 0  ~,\\
\frac{\braket{D_1(v',\epsilon)|V^\mu |B(v) }}{\sqrt{M_D M_B}} 
& = \frac{\tau(w)}{\sqrt{6}}\left\{(1-w^2)\epsilon^{*\mu}-
\left[3v^\mu + (2-w)v'^\mu \right]\epsilon^*\cdot v
\right\}  ~,\\
\frac{\braket{D_1(v',\epsilon)|A^\mu |B(v) }}{\sqrt{M_D M_B}} 
& = -\frac{i\tau(w)}{\sqrt{6}}(1+w)\varepsilon^{\mu\nu\rho\sigma}
\epsilon^*_\nu v_\rho v'_\sigma ~,\\
\frac{\braket{D_1(v',\epsilon)|T^{\mu\eta} |B(v) }}{\sqrt{M_D M_B}} 
& = -\frac{i\tau(w)}{\sqrt{6}}\left[(1+w)(v-v')^{[\mu}\epsilon^{*\eta]}
+3 v^{[\eta}v'^{\mu]} \epsilon^*\cdot v\right]~.
\end{align}
Similarly, for $D_2^*$ we find
\begin{align}
\frac{\braket{D_2^*(v',\epsilon)|S |B(v) }}{\sqrt{M_D M_B}} 
& =0 ~,\\
\frac{\braket{D_2^*(v',\epsilon)|P |B(v) }}{\sqrt{M_D M_B}} 
& = \tau(w) \epsilon^*_{\mu\nu} v^\mu v^\nu ~,\\
\frac{\braket{D_2^*(v',\epsilon)|V^\mu |B(v) }}{\sqrt{M_D M_B}} 
& = -i\tau(w) \varepsilon^{\mu\nu\rho\sigma} \epsilon^*_{\nu\alpha} v^\alpha v_\rho v'_\sigma ~,\\
\frac{\braket{D_2^*(v',\epsilon)|A^\mu |B(v) }}{\sqrt{M_D M_B}} 
& = -\tau(w)  \epsilon^*_{\nu\alpha} v^\alpha \left[(1+w) g^{\mu\nu}-v^\nu v'^\mu\right]~,\\
\frac{\braket{D_2^*(v',\epsilon)|T^{\mu\eta} |B(v) }}{\sqrt{M_D M_B}} 
& = -\tau(w)  \varepsilon^{\mu\eta\rho\sigma}\epsilon^*_{\rho\alpha}v^\alpha(v+v')_\sigma~.
\end{align}
Above $v,\,v'$ are, respectively, the velocities of the $B,D^{**}$
mesons, $A^{[\mu},B^{\nu]}$ stands for the anti-symmetrization $A^\mu
B^\nu - A^\nu B^\mu$, $\varepsilon^{\alpha\beta\mu\nu}$ is the
completely anti-symmetric tensor, and $\epsilon^\mu$ and
$\epsilon^{\mu\nu}$ are spin-one and spin-two polarization
tensors. For the $D_2^*$ meson moving in the $\hat{z}$ direction, we
use the massive spin two polarization tensors
\begin{align}
\epsilon^{\mu\nu}(0) &= \frac{1}{\sqrt{6}}\begin{pmatrix}
-\frac{2p^2}{m^2}&0&0&-\frac{2p E}{m^2}\\0&1&0&0\\0&0&1&0\\-\frac{2p E}{m^2}&0&0&-\frac{2E^2}{m^2}
\end{pmatrix} ~,\\[7pt]
\epsilon^{\mu\nu}(\pm 1) &= \frac{1}{2}\begin{pmatrix}
0&\frac{p}{m}&\mp i\frac{p}{m}&0\\\frac{p}{m}&0&0&\frac{E}{m}\\
\mp i\frac{p}{m}&0&0&\mp i\frac{E}{m}\\0&\frac{E}{m}&\mp i\frac{E}{m}&0
\end{pmatrix} ~, \\[7pt]
\epsilon^{\mu\nu}(\pm 2) &= \frac{1}{2}\begin{pmatrix}
0&0&0&0\\0&1&\mp i&0\\0&\mp i&-1&0\\0&0&0&0
\end{pmatrix}
~.
\end{align}

Finally, for the Isgur-Wise function $\tau(\omega)$ 
we follow the leading order result of~\cite{Bernlochner:2017jxt}, ${\tau(w)\simeq 2-0.9\,\omega}$. We checked that our results are insensitive to
$\mathcal{O}(1)$ modification of this function. 

\section{Modeling the $D^{**}$ decay}\label{app:D** to D*pi decay}
In order to model the $D^{**}$ decay, we relate helicity amplitudes to
partial waves. Using the helicity formalism~ ($e.g.$
\cite{Richman:1984gh,Haber:1994pe}) the $D^{**}\to D^* \pi$ amplitude
is given by
\begin{align}\label{eq:D** decay amplitude}
\mathcal{A}(D^{**}(\lambda)\to D^{*+}(\lambda')\pi^-) = \sqrt{\frac{2J+1}{4\pi}}
e^{-i\varphi(\lambda-\lambda')}
d^J_{\lambda\lambda'}(\theta_D)	\mathcal{A}^J_{\lambda' 0}~,
\end{align}
where $J$ is the spin of the $D^{**}$ meson, $\lambda$ and $\lambda'$ 
are the helicities of the $D^{**}$ and $D^*$ mesons, respectively,
and $d^J_{\lambda\lambda'}$ are the Wigner
functions~\cite{Patrignani:2016xqp}. In the helicity amplitude $\mathcal{A}^J_{\lambda'0}$ the zero stands for the pion helicity. In the case of $D^{**}$ mesons, the projection of the helicity amplitudes to partial waves is given by~\cite{Jacob:1959at}
\begin{align}
\mathcal{A}^{J}_{\lambda' 0}  & =
(-1)^{1-J} \sqrt{\frac{1}{2J+1}} C_{10}(J,\lambda';\lambda',0) S + \sqrt{\frac{5}{2J+1}} C_{21}(J,\lambda';0,\lambda') D ~,
\end{align}
where $S,\,D$ are partial wave functions, and $C_{j_1 j_2} (J,M;m_1,m_2)$ are Clebsch-Gordan coefficients.
It follows that
\begin{align}
\mathcal{A}^1_{0 0} &  =\sqrt{\frac{1}{3}} S - \sqrt{\frac{2}{3}}  D~ , \quad
\mathcal{A}^1_{1 0}  = \mathcal{A}^1_{-1 0}  = \sqrt{\frac{1}{3}} S + \sqrt{\frac{1}{6}}  D~, \label{eqApp:D1 amplitudes}\\
\mathcal{A}^2_{0 0} &  = 0~, \quad
\mathcal{A}^2_{1 0}   = - \mathcal{A}^2_{-1 0} = -\sqrt{\frac{1}{2}}D~.
\end{align}
We emphasize that the above results are exact. In order to proceed, we
use leading order HQET for $D^{**}$ decays~\cite{Lu:1991px}. To
leading order $D_1$ does not decay through S-wave, thus
$\mathcal{A}^1_{0 0} = -2 \mathcal{A}^1_{1 0}$. Another prediction of
leading order HQET is that the ratio of the decay rates of $D_1 \to
D^*\pi$ and  $D_2^* \to D^*\pi$ is $5/3$. We,
therefore, find
\begin{align}
\frac{\Gamma(D_1 \to D^* \pi)}{\Gamma(D_2^* \to D^* \pi)} & \simeq
\frac{|\mathcal{A}^1_{0 0}|^2 + 2 |\mathcal{A}^1_{1 0}|^2}{2|\mathcal{A}^2_{1 0}|^2} \Rightarrow \frac{|\mathcal{A}^1_{1 0}|^2}{|\mathcal{A}^2_{1 0}|^2} = \frac{5}{9}~.
\end{align}

\bibliography{ref}
\end{document}